\documentclass[10pt,aps,pra,twocolumn,superscriptaddress,nofootinbib]{revtex4-2}
\usepackage{amsmath,amssymb,amsfonts}
\usepackage{braket}
\usepackage{multirow}
\usepackage{graphicx}
\usepackage{hyperref}
\usepackage{physics}
\usepackage{xcolor}
\usepackage{float}
\usepackage{tikz}
\hypersetup{
    colorlinks = true,
    linkcolor = blue,
    citecolor = blue,
    urlcolor = blue
}
\usepackage{booktabs}
\begin{document}
\title{Variational Probe and Measurement Optimization for Structured Phase Estimation}
\author{Priyam Srivastava}
\affiliation{Department of Informatics and Networked Systems, University of Pittsburgh}
\author{Vivek Kumar}
\affiliation{Department of Informatics and Networked Systems, University of Pittsburgh}
\author{Gurudev Dutt}
\affiliation{Department of Physics and Astronomy, University of Pittsburgh}
\author{Kaushik P.~Seshadreesan}
\affiliation{Department of Informatics and Networked Systems, University of Pittsburgh}
\affiliation{Department of Physics and Astronomy, University of Pittsburgh}

\begin{abstract}
We present a proof-of-principle study of variational quantum sensing for estimating a structured linear function of local phase parameters, in which each qubit in a spin-$1/2$ array accumulates a phase $\phi_i = \alpha_i\theta$ with known weights $\vec{\alpha}$ and a global parameter $\theta$. In a hardware-motivated regime of shallow circuits and shallow decoding measurements, we optimize the probe state with respect to the classical Fisher information (CFI) using the Covariance Matrix Adaptation Evolution Strategy. The variational ansatz is built from dipolar-interacting gates and global rotations on a polygon-centered qubit layout. To assess whether the standard Ramsey readout extracts all available information, we introduce a shallow global decoder and optimize it independently with the encoder frozen. For uniform ($\alpha_i = 1/N$) and weighted-central ($\alpha_c = 1$, $\alpha_i = 0.5$) encodings with $N = 2$--$8$ qubits and depths $L = 1$--$3$, the optimized probes approach the respective entanglement-enhanced precision bounds, which reduce to the Heisenberg limit only for uniform encoding. The decoder provides systematic but modest CFI gains. For uniform encoding, these gains are smallest at the deepest circuits, confirming that fixed Ramsey readout is near-optimal for well-converged probes. For weighted encoding, a persistent component remains, reflecting the broken permutation symmetry of the generator under unequal weights. At large $N$, the weighted-encoding CFI also exhibits non-monotonic growth with system size, revealing an expressivity limit of the polygon-symmetric ansatz under asymmetric encoding.
\end{abstract}
\maketitle

%
%
%
\section{Introduction}
Quantum sensors, consisting of spatially distributed probes that share quantum correlations, can achieve sensitivities beyond what is possible with independent probes~\cite{Zhang_2021,Pezzè2021}. In quantum metrology~\cite{PhysRevLett.96.010401,Giovannetti2011}, the ultimate precision attainable by a probe state is bounded by its quantum Fisher information (QFI), while the precision realized by a specific measurement is captured by the classical Fisher information (CFI). For unentangled probes the QFI scales linearly with the number of probes $N$, giving rise to the standard quantum limit (SQL), while entangled probes can achieve quadratic scaling, the Heisenberg limit (HL). How closely these limits can be approached depends on the ability to generate entanglement, on the measurement strategy employed, and on how the parameter of interest is encoded across the probes~\cite{Sidhu2020-nc}. For uniform phase encoding in the ideal symmetric Ramsey setting, the CFI can saturate the QFI, recovering the HL for optimal probe states.

In practice, different probes in a sensor array may couple to the target field with unequal strengths, due to differences in spatial position, interrogation time, or coupling geometry. Such settings arise in distributed field sensing~\cite{PhysRevLett.120.080501}, synchronized clock networks~\cite{Giovannetti2001-qc}, and biomedical applications~\cite{Aslam2023}. The resulting estimation task can be cast as a single-parameter problem, in which each probe acquires a phase $\phi_i = \alpha_i\,\theta$, where $\vec{\alpha} = (\alpha_1, \ldots, \alpha_N)$ is a known weight vector encoding the coupling structure and $\theta$ is the global parameter of interest. The attainable precision depends on $\vec{\alpha}$, interpolating between the SQL and an entanglement-enhanced (EE) bound that departs from the HL for non-uniform encodings~\cite{PhysRevA.97.042337}. For permutation-symmetric encodings, GHZ states are the optimal probe and achieve the HL under standard Ramsey readout.

Although GHZ-like probe states can, in principle, saturate these bounds, preparing large entangled states on near-term hardware remains challenging. For dipolar-coupled solid-state spins, early demonstrations of entanglement between separate defect electron spins achieved a Bell-state fidelity of $0.67$~\cite{Dolde2013}, and scaling such entanglement to multi-qubit registers remains an active experimental challenge. More broadly, GHZ-state fidelity on solid-state spin platforms degrades with qubit count. Even in state-of-the-art silicon processors with per-gate fidelities exceeding $0.99$, the GHZ fidelity falls from ${\sim}0.9$ at three qubits to near the entanglement threshold at eight~\cite{Edlbauer2025}. Across platforms, the number of entangling operations required for larger GHZ states compounds gate and coherence errors, so reliable state preparation in the near term is confined to shallow circuits built from hardware-native interactions.

These limitations define a constrained regime for quantum sensing. The accessible probe states are restricted to those producible by shallow hardware-native circuits, and the accessible measurements are correspondingly limited to those generated by shallow decoding circuits before a fixed computational-basis readout. In the unconstrained setting, with access to the full Hilbert space and to arbitrary measurements, the optimal probe and measurement follow directly from the generator spectrum. Under the joint constraints considered here, neither the optimal finite-depth probe state nor the corresponding optimal measurement is specified by the generator alone, even for a single-parameter problem with a known generator. For non-uniform encodings, the asymmetric generator compounds the difficulty further, and the optimal measurement basis for the resulting probe state is not known a priori.

In this constrained regime, variational quantum algorithms offer a constructive route to identifying probe and measurement strategies without requiring analytic knowledge of the optimal state. Zheng et al.~\cite{Zheng2022-bt} optimized dipolar-interacting variational circuits to approach the HL for uniform encoding, treating the measurement as a fixed Ramsey readout. In a related line of work, Kaubruegger et al.~\cite{PhysRevX.11.041045} and Marciniak et al.~\cite{Marciniak2022} developed Bayesian variational schemes that jointly optimize the probe state and an inverse-encoder decoding unitary, demonstrating close-to-optimal Ramsey interferometry on trapped ions with up to 26 qubits. These works, however, operate in the permutation-symmetric setting with uniform phase encoding and platform-native interactions, where the generator spectrum and symmetry largely dictate the optimal state. Meyer~\cite{Meyer2021-zl} introduced a general variational toolbox for joint probe-measurement optimization applicable, in principle, to arbitrary encodings, though the demonstrations focus on abstract circuit ansätze and do not address structured linear functions under hardware-native interactions. The questions of whether a hardware-motivated ansatz can prepare the asymmetric entanglement demanded by a non-uniform generator, and whether the standard Ramsey readout remains adequate once permutation symmetry is broken, have not been investigated in any of these frameworks.

In this work, we present a proof-of-principle study of variational probe optimization for estimating structured linear functions under noiseless evolution. Using parameterized quantum circuits composed of dipolar-interacting gates and global rotations, we optimize entangled probe states with respect to the CFI. The optimizer is the Covariance Matrix Adaptation Evolution Strategy (CMA-ES), a gradient-free evolutionary method run with multiple independent seeds. Circuit depths are restricted to $L = 1$--$3$ layers, consistent with the hardware motivation above, even though our simulations are noiseless. To assess whether the standard Ramsey readout fully extracts the information available in the variationally prepared probes, we introduce a shallow global decoder and optimize it independently with the encoder frozen. Any CFI improvement is therefore attributable solely to measurement-basis adjustment, giving the decoder a clean diagnostic role.

Two encoding profiles are investigated on a polygon-centered qubit layout, namely uniform ($\alpha_i = 1/N$) and weighted-central ($\alpha_c = 1$, $\alpha_i = 0.5$). For $N = 2$--$8$ qubits and depths $L = 1$--$3$, the optimized probes approach the respective EE precision bounds~\cite{PhysRevA.97.042337}~(Appendix~\ref{app:bounds}). The decoder provides systematic but modest CFI gains, which separate into two effects. Convergence compensation diminishes with depth as the encoder improves, while structural basis misalignment under asymmetric encoding persists even for well-converged probes. In addition, at large $N$ the weighted-encoding CFI fails to sustain monotonic growth across all depths and seed runs, revealing a structural mismatch between the polygon-symmetric ansatz and the asymmetric entanglement required by the weighted generator.

Section~\ref{sec:framework} introduces the variational framework. Section~\ref{sec:results} presents numerical results and their physical interpretation. Section~\ref{sec:conclusion} summarizes findings and outlines extensions.

\begin{figure*}[!htbp]
  \centering
  \includegraphics[width=1.0\textwidth]{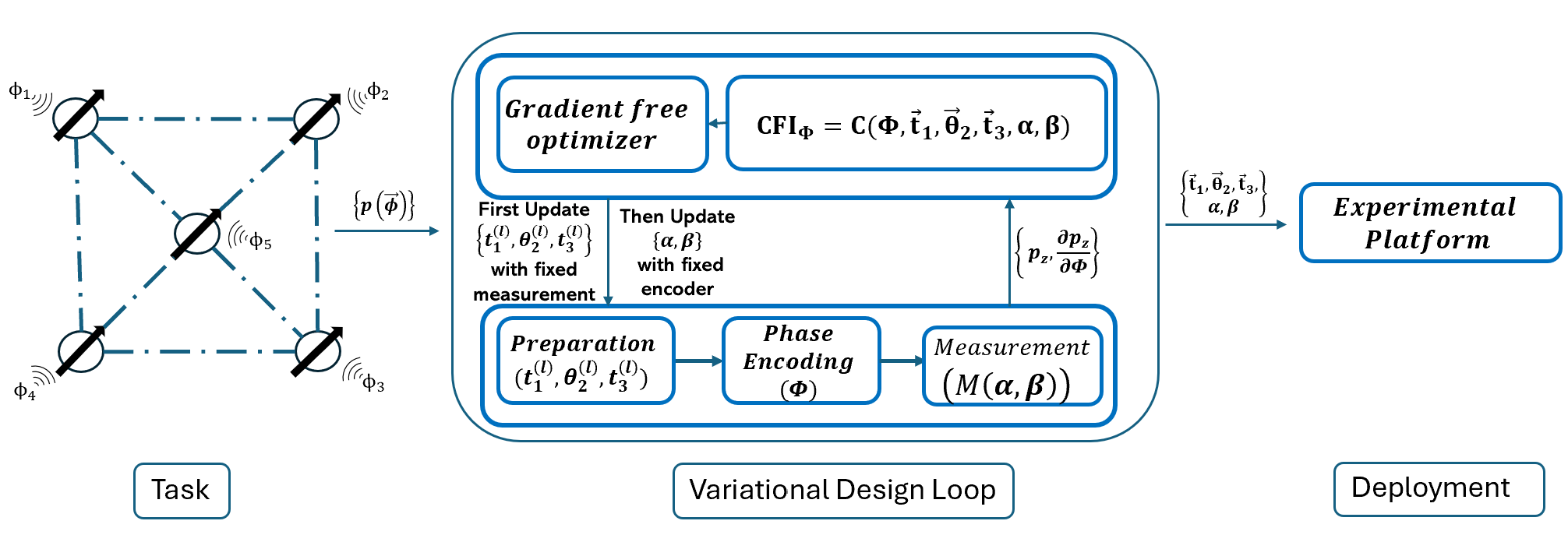}
  \caption{Overview of the variational quantum sensing protocol.
  (\textbf{Task}) A structured phase profile, defined by a weight vector
  $\vec{\alpha}$, is imprinted across the qubits.
  (\textbf{Variational design}) A hardware-efficient ansatz with dipolar
  interactions and collective rotations is optimized to minimize a
  metrological cost function, here based on the Classical Fisher
  Information (CFI).
  (\textbf{Deployment}) The optimized probe can then be implemented on an
  experimental platform for enhanced parameter estimation.}
  \label{fig:protocol}
\end{figure*}

%
%
\section{Structured Encoding and Variational Framework}
\label{sec:framework}

As illustrated in Fig.~\ref{fig:protocol}, variational quantum sensing can be viewed as a three-stage workflow. A structured phase profile,
defined by a weight vector~$\vec{\alpha}$, is imprinted across the qubits according to the chosen encoding. A variational design loop then prepares candidate probe states and optimizes them against a
metrological cost function, producing circuit parameters adapted to the task. The resulting probes can, in principle, be deployed on experimental hardware.

At the core of this approach are Variational Quantum Algorithms (VQAs)~\cite{Cerezo2021-hy}. A parameterized quantum circuit acts on a
simple initial state to generate correlated probes, which are evaluated using the cost function. A classical-quantum feedback loop updates the
circuit parameters, gradually improving sensitivity and yielding probe states that approach entanglement-enhanced precision under the specified encoding. The adequacy of the readout is assessed independently through a shallow global decoder, as described in
Sec.~\ref{subsec:decoder}.

The remainder of this section introduces the components of our framework. Section~\ref{subsec:encoding} defines the structured phase encoding and the precision bounds relevant to each profile. Section~\ref{subsec:ansatz} describes the layered ansatz constructed from dipolar interactions and collective rotations. Section~\ref{subsec:decoder} introduces the global decoder and its physical justification. Section~\ref{subsec:cfi} defines the CFI and
the directional parameter-shift rule used to evaluate it. Finally, Section~\ref{subsec:optimization} presents the CMA-ES routine and the multi-seed warm-start scheme.

\subsection{Directional Phase Encoding for Structured Parameter Estimation}
\label{subsec:encoding}

To tailor the probe state to a given sensing task, we imprint a directionally 
weighted phase profile across the qubits. Each qubit $i$ accumulates a local phase
\begin{equation}
    \phi_i = \alpha_i \theta,
    \label{eq:directional_phi}
\end{equation}
where $\theta$ is the global parameter to be estimated and 
$\vec{\alpha}=(\alpha_1,\alpha_2,\dots,\alpha_N)$ is a fixed weight vector that 
encodes spatial structure or task-specific sensitivity. Under this directional 
encoding, the estimation problem reduces to a single scalar parameter
\begin{equation}
    q \;=\; \sum_{i=1}^N \alpha_i\,\phi_i \;=\; \|\vec{\alpha}\|^2\,\theta,
    \label{eq:q}
\end{equation}
where the non-trivial content lies in the structure of $\vec{\alpha}$, which shapes 
both the optimal probe state and, as we discuss in Sec.~\ref{subsec:decoder}, the 
optimal measurement basis.

Operationally, the encoding is implemented with local $Z$-rotations,
\begin{equation}
    R_{z}^{(i)}(\phi_i) \;=\; \exp\!\left(-\,\frac{i}{2}\,\alpha_i \theta\, 
    \sigma_z^{(i)}\right),
    \label{eq:RZ_encoding}
\end{equation}
so that the global operation factorizes as
\begin{equation}
    U(\theta) \;=\; \bigotimes_{i=1}^N R_{z}^{(i)}(\alpha_i \theta)
                 \;=\; \exp\!\left(-\,i\,\theta\,\hat{H}_\theta\right),
    \label{eq:encoding}
\end{equation}
with the generator of $\theta$-translations
\begin{equation}
  \hat{H}_\theta \;=\; \frac{1}{2}\sum_{i=1}^{N} \alpha_i\,\sigma_z^{(i)}.
  \label{eq:generator}
\end{equation}

The attainable precision depends directly on the structure of $\vec{\alpha}$. 
Following Ref.~\cite{PhysRevA.97.042337}, the SQL for separable probes and the 
EE bound for optimally entangled probes are
\begin{equation}
    F_\mathrm{SQL}(q) = \frac{1}{\|\vec{\alpha}\|^2}, \qquad
    F_\mathrm{EE}(q)  = \frac{S^2}{\|\vec{\alpha}\|^4},
    \label{eq:bounds}
\end{equation}
where $S = \sum_i |\alpha_i|$. For uniform encoding ($\alpha_i = 1/N$), these 
reduce to the familiar $F_\mathrm{SQL} = N$ and $F_\mathrm{EE} = N^2$, recovering the standard HL. For the weighted-central profile $\vec{\alpha} = (1, 0.5, \dots, 0.5)$, the EE bound is strictly below $N^2$, reflecting the reduced symmetry of 
the generator. Numerical values for both encodings are tabulated in Appendix~\ref{app:bounds}.

We investigate two encoding profiles throughout this work. In the uniform profile ($\alpha_i = 1/N$), all qubits accumulate equal phases, relevant to array magnetometry~\cite{Pham_2011} and synchronized clock networks~\cite{Kómár2014}. In the weighted-central profile ($\alpha_c = 1$, $\alpha_i = 0.5$ for $i \neq c$), the central qubit dominates the encoding, modeling near-field sensing scenarios~\cite{Maletinsky2012}. Uniform encoding
preserves the permutation symmetry of the polygon geometry, while weighted-central encoding breaks it, with direct consequences for both the optimal probe state and the optimal measurement.

\subsection{Variational Ansatz for Probe State Preparation}
\label{subsec:ansatz}

We construct a parameterized quantum circuit to prepare entangled probe states adapted to the structured encodings of Sec.~\ref{subsec:encoding}. The variational ansatz is composed of repeated layers of unitary operations built from dipolar interactions and collective rotations. Each qubit is modeled as a spin-$\tfrac{1}{2}$ particle with vector operator $\vec{S}_i = (S_i^x, S_i^y, S_i^z)$, and the interactions between them are governed by a
dipolar Hamiltonian:
\begin{equation}
\hat{H}_{\mathrm{int}} = \sum_{i<j} V_{ij} \left( J_I\, S_i^z S_j^z 
    + J_S\, \vec{S}_i \cdot \vec{S}_j \right),
\label{eq:dipolar_hamiltonian}
\end{equation}
where the interaction strength $V_{ij}$ depends on spatial coordinates $\vec{r}_i$ and the angle $\beta_{ij}$ between the inter-qubit axis and an external bias field:
\begin{equation}
V_{ij} \;=\; \frac{\mu_{0}\,\gamma^{2}\,\hbar^{2}}{4\pi\,\|\mathbf{r}_i - 
\mathbf{r}_j\|^{3}}\,\Bigl[\,1 \;-\; 3\,\cos^2(\beta_{ij})\Bigr].
\label{eq:Vij}
\end{equation}
This Hamiltonian captures long-range dipolar coupling in systems such as NV centers or Rydberg arrays and is adapted from Ref.~\cite{Zheng2022-bt}. The coupling constants take values $J_I = 3$ and $J_S = -1$, with $\mu_0 = 6.7\times10^{-4}$ and $\gamma = 1.001$.

The coupling strengths $V_{ij}$ are determined by the underlying geometry of the qubit array. In this work, we consider polygon-centered lattice configurations for $N = 2$--$8$, each consisting of a central qubit surrounded symmetrically by $(N{-}1)$ peripheral qubits arranged at equal angular spacing. This layout preserves permutation symmetry within the peripheral shell while distinguishing the central site, making it a natural platform for both the uniform and weighted-central encoding schemes studied here.

\begin{figure}[htbp]
  \centering
  \resizebox{0.9\columnwidth}{!}{%
  \tikzset{
    qubitC/.style={circle, draw=red!70!black,  fill=red!70!black,  inner sep=2pt},
    qubitP/.style={circle, draw=blue!70!black, fill=blue!70!black, inner sep=2pt},
    edge/.style ={draw=gray!70, line width=0.55pt},
    every node/.style={font=\scriptsize}
  }
  \begin{tikzpicture}[scale=1.1]
    \begin{scope}[shift={(-3.0, 1.9)}]
      \node[qubitC] (c2) at (-1.6,0) {};
      \node[qubitP] (p2) at ( 1.6,0) {};
      \draw[edge]   (c2) -- (p2);
      \node at (0,-0.9) {$N=2$};
    \end{scope}
    \begin{scope}[shift={(2.5, 1.9)}]
      \node[qubitP] (pL) at (-1.6,0) {};
      \node[qubitC] (c3)  at ( 0,0)  {};
      \node[qubitP] (pR) at ( 1.6,0) {};
      \draw[edge] (c3) -- (pL) (c3) -- (pR) (pL) -- (pR);
      \node at (0,-0.9) {$N=3$};
    \end{scope}
    \begin{scope}[shift={(-3.3,-2.1)}]
      \node[qubitC] (c4) at (0,0) {};
      \foreach \idx/\ang in {1/90,2/210,3/330}{
        \node[qubitP] (p\idx) at (\ang:1.8) {};
        \draw[edge] (c4) -- (p\idx);
      }
      \draw[edge] (p1) -- (p2) -- (p3) -- (p1);
      \node at (0,-2.35) {$N=4$};
    \end{scope}
    \begin{scope}[shift={(2.5,-2.1)}]
      \node[qubitC] (c5) at (0,0) {};
      \foreach \idx/\ang in {1/45,2/135,3/225,4/315}{
        \node[qubitP] (p\idx) at (\ang:1.8) {};
        \draw[edge] (c5) -- (p\idx);
      }
      \draw[edge] (p1) -- (p2) -- (p3) -- (p4) -- (p1);
      \node at (0,-2.35) {$N=5$};
    \end{scope}
  \end{tikzpicture}}
  \caption{Representative polygon-centered lattice geometries for $N=2$--$5$. Central qubits (red) are surrounded by $(N{-}1)$ peripherals (blue) at equal angular spacing; grey lines indicate dipolar couplings. Configurations for $N=6$--$8$ follow the same construction.}
  \label{fig:lattice}
\end{figure}

Figure~\ref{fig:lattice} illustrates the layouts for $N=2$--$5$ as representative examples; larger registers up to $N=8$ follow the same polygon-centered construction. The variational ansatz is built from layered gates that combine dipolar interactions with global rotations. The circuit begins with a global $R_y(\pi/2)$ rotation that aligns the initial state $\ket{0}^{\otimes N}$ along the 
$x$-axis. Each layer $\ell$ of the circuit takes the form:
\begin{equation}
\begin{aligned}
\hat{U}^{(\ell)}(t_1^{(\ell)}, \theta_{2}^{(\ell)}, t_3^{(\ell)}) = \;
& R_y\!\left(\tfrac{\pi}{2}\right) \exp\!\left(-i t_3^{(\ell)} \hat{H}_{\mathrm{int}}
  \right) R_y\!\left(-\tfrac{\pi}{2}\right) \\
& \times R_x(\theta_{2}^{(\ell)}) \exp\!\left(-i t_1^{(\ell)} 
  \hat{H}_{\mathrm{int}}\right),
\end{aligned}
\label{eq:ansatz_layer}
\end{equation}
introducing three variational parameters $(t_1^{(\ell)}, \theta_{2}^{(\ell)}, t_3^{(\ell)})$ per layer, for a total of $3L$ encoder parameters at depth $L$. The full circuit state is
\begin{equation}
\ket{\psi(\vec{\theta})} = \left[ \prod_{\ell=1}^L \hat{U}^{(\ell)} \right] 
\ket{+}^{\otimes N}.
\label{eq:full_circuit}
\end{equation}
This ansatz flexibly captures a range of entangled states, including spin-squeezed and GHZ-like states relevant to metrology. The evolution $\exp(-i t \hat{H}_{\mathrm{int}})$ is implemented using first-order Trotterization~\cite{Sieberer2019} via qml.ApproxTimeEvolution in PennyLane; convergence of the CFI with respect to the number of Trotter steps is verified in Appendix~\ref{app:trotter}.

The layered ansatz is architecture-agnostic with respect to the encoding: the same gate structure is used for both profiles, with variational parameters trained separately for each $\vec{\alpha}$. The resulting resource state therefore adapts to the task-specific generator $\hat{H}_\theta$ without requiring explicit symmetry constraints. The directional phase is imprinted by local $R_z(\alpha_i\theta)$  rotations applied after the entangling block, as defined in Sec.~\ref{subsec:encoding}. The choice of readout applied after encoding is the subject of the following section.

\subsection{Global Decoder for Measurement Optimization}
\label{subsec:decoder}

In quantum parameter estimation, the CFI obtained from a given measurement is bounded above by the QFI, with equality when the measurement basis
coincides with the eigenbasis of the symmetric logarithmic derivative~\cite{Helstrom1969,BraunstienCaves1994,Paris2009}. The standard Ramsey readout achieves this equality for certain probe states, but variational circuits of finite depth do not generally prepare such states. It is therefore natural to ask whether adapting the measurement basis to the actual probe state can recover some of the lost information. We address this by introducing a variational global decoder.

In standard Ramsey interferometry, the encoded state
$\ket{\psi_\theta}$ is rotated by a global $R_X(\pi/2)$ prior to $Z$-basis measurement,
\begin{equation}
    \ket{\psi'} = R_X\!\left(\frac{\pi}{2}\right)\ket{\psi_\theta},
\end{equation}
which, in the Heisenberg picture, transforms the measurement observable as
\begin{equation}
    R_X^\dagger\!\left(\frac{\pi}{2}\right)\, Z \,
    R_X\!\left(\frac{\pi}{2}\right) \;=\; Y.
    \label{eq:ramsey_observable}
\end{equation}
The standard Ramsey readout, therefore, corresponds to a fixed measurement in the $Y$ eigenbasis.

To generalize this readout, we introduce a variational decoder applied globally to all qubits before the $Z$ measurement. Up to a global phase, any single-qubit unitary admits the Euler decomposition
\begin{equation}
    U(\alpha,\beta,\gamma) = R_Z(\gamma)\,R_X(\beta)\,R_Z(\alpha),
    \label{eq:euler_decoder}
\end{equation}
where in circuit order $R_Z(\alpha)$ acts first on the state, $R_X(\beta)$ second, and $R_Z(\gamma)$ last, immediately before the $Z$-basis projective measurement. The final $R_Z(\gamma)$ rotation is
redundant: in the Heisenberg picture, the effective observable on each qubit is $U^\dagger Z\, U$, and since $R_Z(\gamma) = e^{-i\gamma\sigma_z/2}$ commutes with $Z$, this reduces to
\begin{equation}
    U^\dagger Z\, U = R_Z^\dagger(\alpha)\,R_X^\dagger(\beta)\,
    Z\, R_X(\beta)\,R_Z(\alpha),
    \label{eq:effective_obs}
\end{equation}
independent of $\gamma$. The same holds for every computational-basis projector $\ket{z}\!\bra{z}$, so $\gamma$ does not affect any outcome probability or the resulting CFI.

The three-parameter Euler form therefore reduces to an effective two-parameter decoder
\begin{equation}
    V(\alpha,\beta) = \bigotimes_{i=0}^{N-1}
    R_X(\beta)\,R_Z(\alpha),
    \label{eq:decoder}
\end{equation}
where in circuit order $R_Z(\alpha)$ acts first on the encoded state followed by $R_X(\beta)$ before the $Z$ measurement. This decoder generalizes the standard Ramsey readout, which corresponds to the
special case $\alpha = 0$, $\beta = \pi/2$: the parameter $\alpha$ rotates the measurement axis within the equatorial plane of the Bloch sphere, while $\beta$ controls the polar tilt away from the $Z$ axis. The parameters $(\alpha,\beta)$ are optimized independently with the encoder frozen to maximize the classical Fisher information.

We employ a global decoder rather than a fully local per-qubit decoder. A local decoder would introduce $2N$ additional parameters, significantly increasing the optimization complexity and potentially leading to overfitting at small system sizes. More importantly, a highly expressive decoder could compensate for deficiencies in the encoder itself, obscuring whether CFI gains reflect genuine measurement-basis mismatch or simply a reallocation of optimization effort. In contrast, the global decoder introduces only two parameters independent of $N$, remains compatible with experimental platforms supporting collective rotations~\cite{Marciniak2022}, and is deliberately kept minimal so that any CFI improvement over the fixed Ramsey baseline can be attributed to the measurement basis rather than to encoder compensation. The relative contributions of the two
mechanisms underlying these gains, convergence compensation and structural basis misalignment, are quantified in Sec.~\ref{subsec:discussion}.

\subsection{Classical Fisher Information and Parameter-Shift Rule}
\label{subsec:cfi}

To evaluate the probe's sensitivity to the encoded parameter $\theta$, we compute the CFI from outcome statistics of computational-basis measurements performed after phase encoding and the decoder of Sec.~\ref{subsec:decoder}. For noiseless Ramsey interferometry with commuting generators and $Z$-basis measurement, the CFI is known to saturate the QFI of the probe state~\cite{PhysRevLett.96.010401}; the decoder introduced above allows this saturation to be approached more closely when the fixed Ramsey readout is suboptimal.

Let $\rho_\theta = U(\theta)\,\rho\,U^\dagger(\theta)$ denote the encoded state, where $U(\theta)$ is the encoding unitary of Eq.~\eqref{eq:encoding} and $\rho$ is 
the variationally prepared probe state. Let $\{P_k\}$ denote the set of computational-basis projectors after the readout unitary (either the fixed Ramsey rotation $R_X(\pi/2)$ or the variational decoder $V(\alpha,\beta)$ of 
Eq.~\eqref{eq:decoder}). The corresponding outcome probabilities are
\begin{equation}
    p_k(\theta) \;=\; \mathrm{Tr}\!\left[P_k\,\rho_\theta\right],
\end{equation}
which, since $\vec{\phi} = \vec{\alpha}\,\theta$, depend only on the scalar $q = \|\vec{\alpha}\|^2\theta$ defined in Eq.~\eqref{eq:q}. We therefore write $p_k = p_k(q)$ and treat the task as single-parameter estimation in $q$.

The CFI with respect to $q$ is then
\begin{equation}
    F(q) \;=\; \sum_k \frac{1}{p_k(q)}
    \!\left(\frac{dp_k}{dq}\right)^{\!2},
    \label{eq:cfi_q}
\end{equation}
with
\begin{equation}
    \frac{dp_k}{dq}
    \;=\; \frac{1}{\|\vec{\alpha}\|^2}
    \sum_{i=1}^{N} \alpha_i\,\frac{\partial p_k}{\partial \phi_i},
    \label{eq:chain_rule}
\end{equation}
where the partial derivatives $\partial p_k/\partial \phi_i$ are obtained via the 
exact parameter-shift rule~\cite{Schuld2019} for Pauli-generated $R_Z$ gates using a shift $s = \pi/2$:
\begin{equation}
    \frac{\partial p_k}{\partial \phi_i}
    \;=\; \frac{p_k(\phi_i{+}s) - p_k(\phi_i{-}s)}{2},
    \qquad s = \frac{\pi}{2}.
    \label{eq:param_shift}
\end{equation}
Substituting Eq.~\eqref{eq:chain_rule} into Eq.~\eqref{eq:cfi_q} yields the 
working expression
\begin{equation}
    F(q) \;=\; \sum_k \frac{1}{p_k}\!\left(
    \frac{1}{\|\vec{\alpha}\|^2}
    \sum_{i=1}^{N} \alpha_i\,\frac{\partial p_k}{\partial \phi_i}
    \right)^{\!2}.
    \label{eq:cfi_final}
\end{equation}

For completeness, the Fisher information with respect to $\theta$ is related by  
$F_\theta = \left(\tfrac{dq}{d\theta}\right)^{\!2} F(q) = \|\vec{\alpha}\|^{4}F(q)$. 
Throughout this work, we report $F(q)$ and compare it with the analytic precision bounds given in Appendix~\ref{app:bounds}. Maximizing $F(q)$ defines the variational objective, directing the circuit parameters toward probe states whose measurement 
statistics are maximally sensitive to the structured parameter $\theta$ specified by $\vec{\alpha}$.

\subsection{CMA-ES Optimization Strategy}
\label{subsec:optimization}

Although gradients of the CFI can, in principle, be computed, we adopt a gradient-free strategy for robustness to rugged landscapes and measurement noise. Specifically, we use CMA-ES, an evolutionary algorithm that adapts a multivariate Gaussian search distribution over circuit parameters and is well-suited to nonconvex, ill-conditioned
objectives~\cite{Hansen2006,Nomura2024-sm}.

It maintains a mean vector and a covariance matrix over the parameter space. At each generation, it samples a population of candidate parameter vectors $\{\vec{\theta}^{(j)}\}$, evaluates a fitness value for each candidate, and updates the mean and covariance to bias future samples toward regions of higher fitness. The parameter dependence of $F(q)$ makes CMA-ES particularly attractive, as it does
not rely on circuit-parameter gradients and thus avoids issues with barren plateaus or noisy gradient estimates.

To compute the directional derivative entering $F(q)$, we use the exact parameter-shift rule for the local $R_z$ encoders (Sec.~\ref{subsec:cfi}). For each candidate $\vec{\theta}^{(j)}$, this
requires evaluating the circuit for the unshifted configuration and for two shifted values of each local phase $\phi_i$ (at $\pm\pi/2$), yielding $(2N{+}1)$ forward evaluations per fitness call. With
first-order Trotterization of the interaction blocks, the per-candidate cost scales as $\mathcal{O}\!\big((2N{+}1)\,L\,m\,|\mathcal{E}|\big)$, where $L$ is the circuit depth, $m$ is the number of Trotter slices per evolution block, and $|\mathcal{E}|$ is the number of dipolar pairs
included by the geometry. In practice, these evaluations are batched to exploit parallelism.

The optimization proceeds in two stages. In the first stage, the encoder parameters $\vec{\theta} = \{t_1^{(\ell)}, \theta_2^{(\ell)}, t_3^{(\ell)}\}_{\ell=1}^L$ are optimized with the
fixed Ramsey readout $R_X(\pi/2)$ held as the measurement. For each $(N, L, \text{encoding})$ configuration, we run 5 independent CMA-ES instances with well-separated seeds and retain the best result.
Multiple independent seeds are essential at large $N$, where the CFI landscape becomes increasingly multimodal. In the second stage, the encoder parameters are frozen, and the decoder parameters
$(\alpha, \beta)$ of Eq.~\eqref{eq:decoder} are optimized independently via CMA-ES, initialized from the fixed Ramsey values $(\alpha, \beta) = (0, \pi/2)$. If no decoder seed improves upon the
fixed Ramsey baseline, the fixed Ramsey parameters are retained, ensuring the reported CFI is never below the fixed-readout result. This two-stage structure treats the decoder as a diagnostic, since any CFI improvement over the fixed Ramsey baseline quantifies the information left on the table by the standard readout for the given probe state.
The sequential design also preserves interpretability, because in a joint optimization, the encoder and decoder parameters can trade off
against each other, making it difficult to attribute CFI gains to either the probe state or the measurement individually.

To accelerate convergence with increasing depth, the encoder optimization employs a layerwise warm start. Depth $L=1$ is optimized from a broad initial distribution; the best parameters then seed depth
$L{+}1$ by appending a new layer initialized near zero while keeping previously learned parameters intact, following the layerwise strategy of Skolik \emph{et al.}~\cite{Skolik2021}. This incremental scheme preserves useful correlations learned at lower depth and limits the search to the newly introduced degrees of freedom, improving stability
and reducing evaluation budget. Standard CMA-ES termination criteria are used (stagnation in fitness and step-size adaptation). Full hyperparameter settings, seed values, and optimization variance data
are given in Appendix~\ref{app:cmaes}.

%
%
%
%
\section{Results and Discussion}
\label{sec:results}

We benchmark the variational sensing framework on both encoding profiles for $N = 2$--$8$ qubits and circuit depths $L = 1$--$3$, using the sequential optimization pipeline of Sec.~\ref{subsec:optimization}. Performance is quantified by the CFI relative to the SQL and the encoding-specific EE bound. Section~\ref{subsec:sequential} reports the 
optimized CFI for both encodings, including the encoder-only baseline under fixed Ramsey readout and the gains from the independently optimized decoder. Section~\ref{subsec:fidelity} presents a GHZ fidelity analysis that characterizes the convergence of the variational encoder toward the
expected optimal probe states. Section~\ref{subsec:discussion} interprets 
the decoder contributions within a two-effect physical framework and discusses the implications for measurement optimality. Seed variance data and Trotterization convergence tests are collected in Appendices~\ref{app:trotter}--\ref{app:seeds}.

%
%
%
%
%
%
\subsection{Encoder Optimization and Decoder Gains}
\label{subsec:sequential}

The encoder is first optimized under fixed Ramsey readout, and the decoder parameters are then optimized independently with the encoder frozen (Sec.~\ref{subsec:optimization}). Complete numerical results for all configurations are provided in the Supplemental Material~\cite{supplemental}.

\begin{figure}[htbp]
    \centering
    \includegraphics[width=\columnwidth]{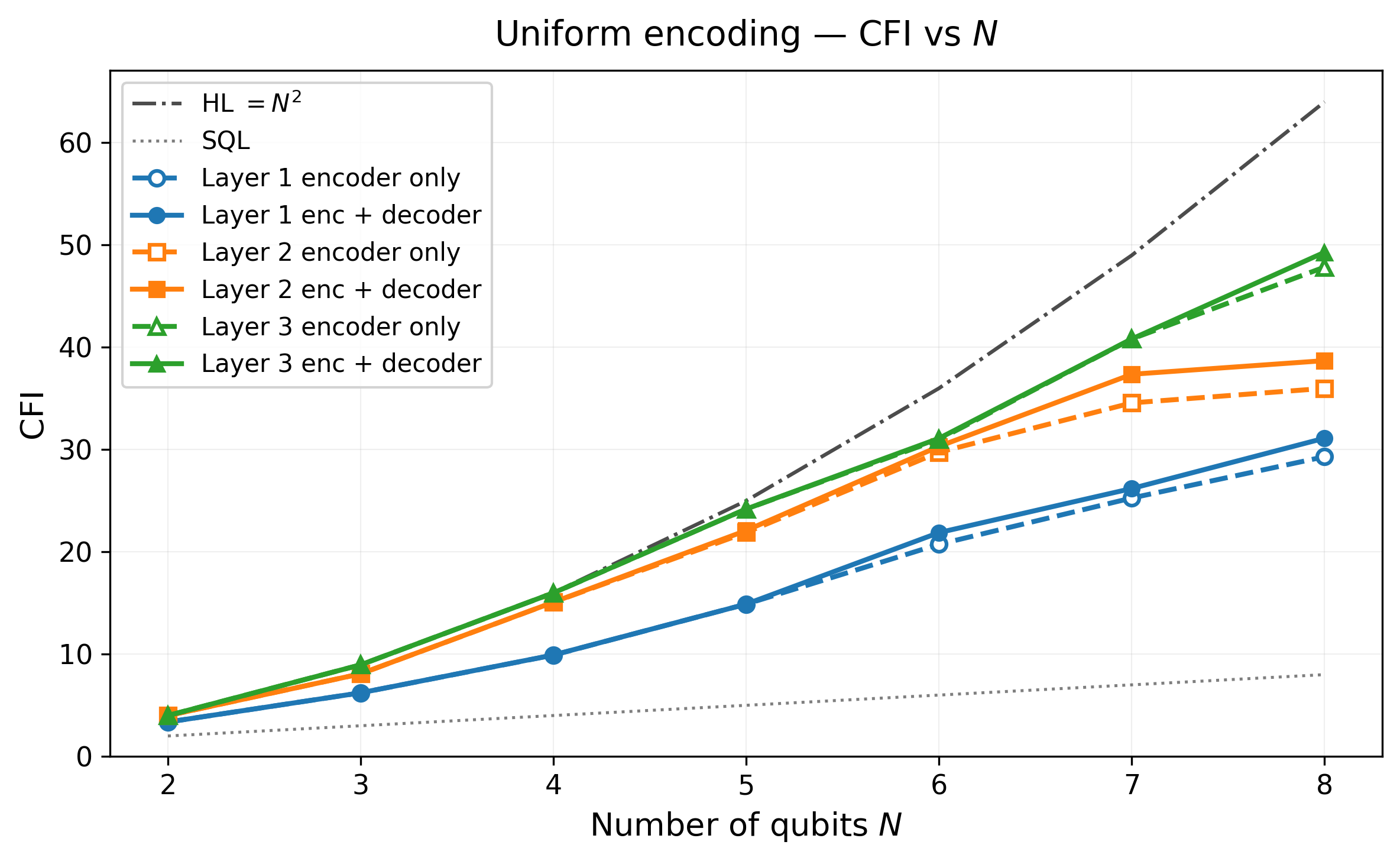}
    \caption{%
    Uniform encoding ($\alpha_i = 1/N$). CFI versus $N$ for $L = 1$, $2$,
    and $3$. Dashed curves: encoder only (fixed Ramsey readout). Solid curves:
    encoder with independently optimized decoder. The Heisenberg limit (HL) and
    SQL are shown for reference.
    }
    \label{fig:seq_uniform}
\end{figure}

\paragraph*{Uniform encoding.}
Figure~\ref{fig:seq_uniform} shows the CFI versus $N$ for $\alpha_i = 1/N$ at depths $L = 1$, $2$, and $3$. Dashed curves give the encoder-only (fixed Ramsey) baseline. Solid curves include the independently optimized decoder. The encoder-only CFI grows monotonically with $N$ at all depths, approaching the HL progressively as circuit depth increases. At $L = 3$, the encoder saturates the HL for $N = 2$ ($100\%$) and nearly saturates it for $N = 3$--$4$ ($99.5\%$ and $99.9\%$, respectively). For $N = 5$ the CFI reaches $96.8\%$ of the HL at $L = 3$, up from $59.6\%$ at $L = 1$. At the largest system sizes, the gap to the HL narrows with each additional layer but remains appreciable, with $N = 7$ reaching $83.3\%$ and $N = 8$ reaching $74.8\%$ at $L = 3$.

Adding the decoder (solid curves) preserves the monotonic trend observed under uniform encoding and, by construction of the fixed Ramsey fallback, never decreases the CFI. Decoder gains are negligible for $N \leq 4$ at all depths. For larger $N$, the gains grow with system size and are smallest at the deepest circuit, though not monotonically. At $N = 8$, the relative improvement is $6.2\%$ at $L = 1$, $7.5\%$ at $L = 2$, and $3.0\%$ at $L = 3$.

\begin{figure}[htbp]
    \centering
    \includegraphics[width=\columnwidth]{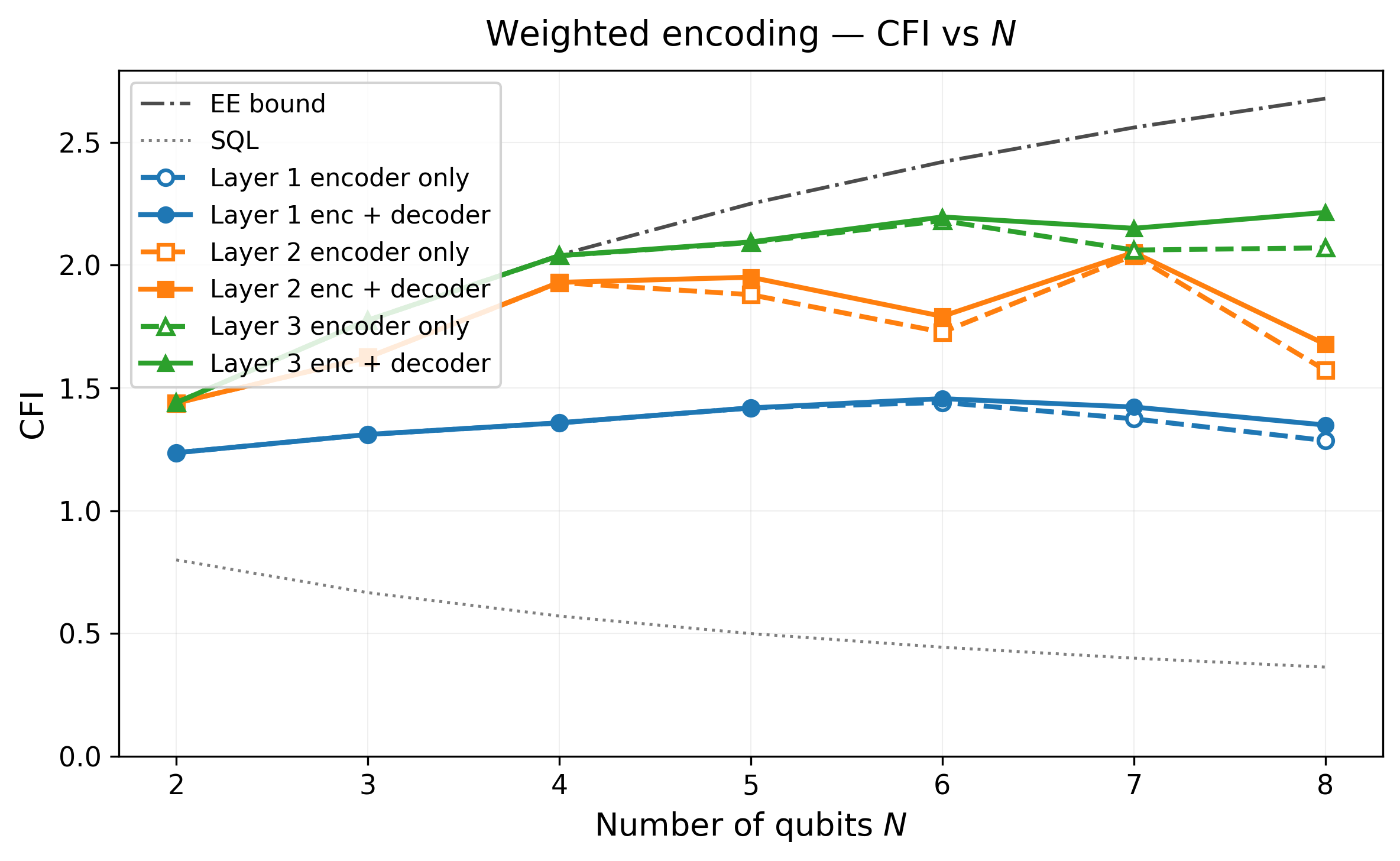}
    \caption{%
    Weighted-central encoding $\vec{\alpha} = (1, 0.5, \ldots, 0.5)$. CFI versus $N$ for $L = 1$, $2$, and $3$. Dashed curves: encoder only (fixed Ramsey readout). Solid curves: encoder with independently optimized decoder. The EE bound and SQL are shown for reference.
    }
    \label{fig:seq_weighted}
\end{figure}

\paragraph*{Weighted-central encoding.}
Figure~\ref{fig:seq_weighted} shows the analogous results for $\vec{\alpha} = (1, 0.5, \ldots, 0.5)$. The encoder-only CFI at $L = 3$ reaches $100\%$ of the EE bound for $N = 2$ and $99.9\%$ for $N = 3$--$4$. For intermediate sizes, $N = 5$ reaches $93.0\%$ and $N = 6$ reaches $90.1\%$ at $L = 3$, both substantial improvements over their $L = 1$ values. However, the CFI turns over at $N = 7$--$8$, where the encoder-only values at $L = 3$ reach only $80.5\%$ and $77.3\%$ of the EE bound. In contrast to the uniform case, decoder gains persist across depths rather than diminishing. At $N = 8$, the relative improvement is $4.9\%$ at $L = 1$, $6.7\%$ at $L = 2$, and $7.0\%$ at $L = 3$. The physical origin of these encoding-dependent decoder gains and the large-$N$ turnover under weighted encoding are discussed in Sec.~\ref{subsec:discussion}.

%
%
%
%
\subsection{GHZ Fidelity Analysis}
\label{subsec:fidelity}

To characterize the extent to which the optimized encoder states acquire GHZ-like structure, we compute the fidelity~\cite{Jozsa01121994} of each
optimized encoder state $\ket{\psi(\vec{\theta})}$ with respect to the $N$-qubit GHZ state
$\ket{\mathrm{GHZ}} = (\ket{0}^{\otimes N} + \ket{1}^{\otimes N})/\sqrt{2}$:
\begin{equation}
    \mathcal{F} = \left|\!\left\langle \mathrm{GHZ} \middle|
    \psi(\vec{\theta}) \right\rangle\!\right|^2.
    \label{eq:fidelity}
\end{equation}
For uniform encoding, the GHZ state is the natural reference because it maximizes the QFI and achieves the HL. For weighted encoding, the optimal probe generally deviates from GHZ because the asymmetric generator breaks permutation symmetry, so the GHZ fidelity should be interpreted there as a diagnostic of GHZ-like character rather than as a certification of exact optimality.
\begin{figure}[htbp]
    \centering
    \includegraphics[width=\columnwidth]{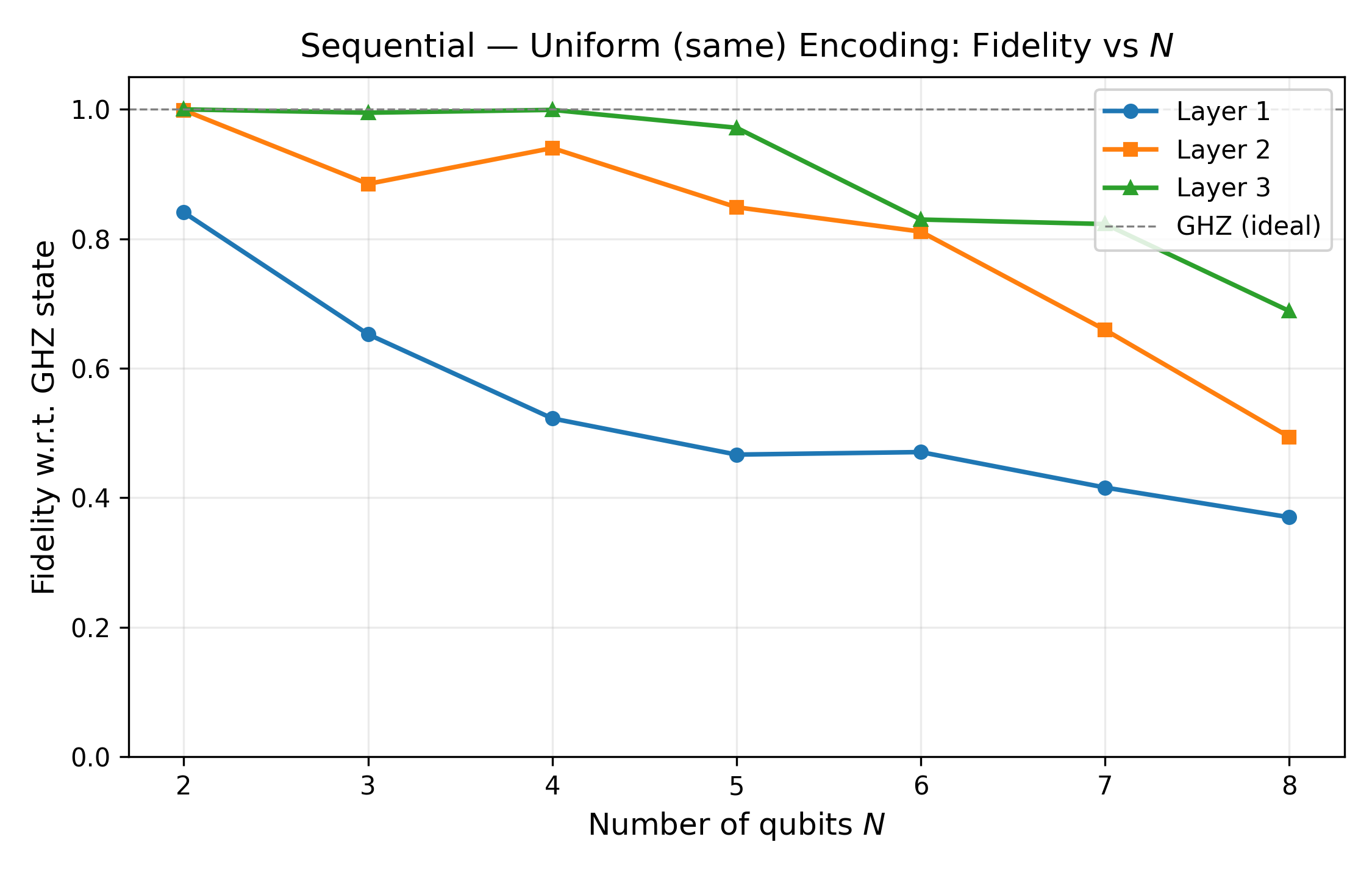}
    \caption{%
    GHZ fidelity versus $N$ for the sequential encoder, uniform encoding
    ($\alpha_i = 1/N$), at depths $L = 1$, $2$, and $3$. The dashed line
    marks ideal fidelity of $1$.
    }
    \label{fig:fid_uniform}
\end{figure}
\paragraph*{Uniform encoding.}
Figure~\ref{fig:fid_uniform} shows the GHZ fidelity versus $N$ for $L = 1$, $2$, and $3$. Fidelity increases monotonically with depth at every system size. At $L = 3$, the encoder reaches fidelities above $0.97$ for $N = 2$--$5$, with $N = 2$--$4$ essentially at unity. For larger systems, the fidelity decreases with $N$ but continues to improve
with each additional layer: at $N = 6$ the fidelity rises from $0.47$ at $L = 1$ to $0.83$ at $L = 3$; at $N = 8$, from $0.37$ to $0.69$. These trends are consistent with the CFI results of  Sec.~\ref{subsec:sequential}; the connection between fidelity and decoder gains is discussed in
Sec.~\ref{subsec:discussion}.
\begin{figure}[htbp]
    \centering
    \includegraphics[width=\columnwidth]{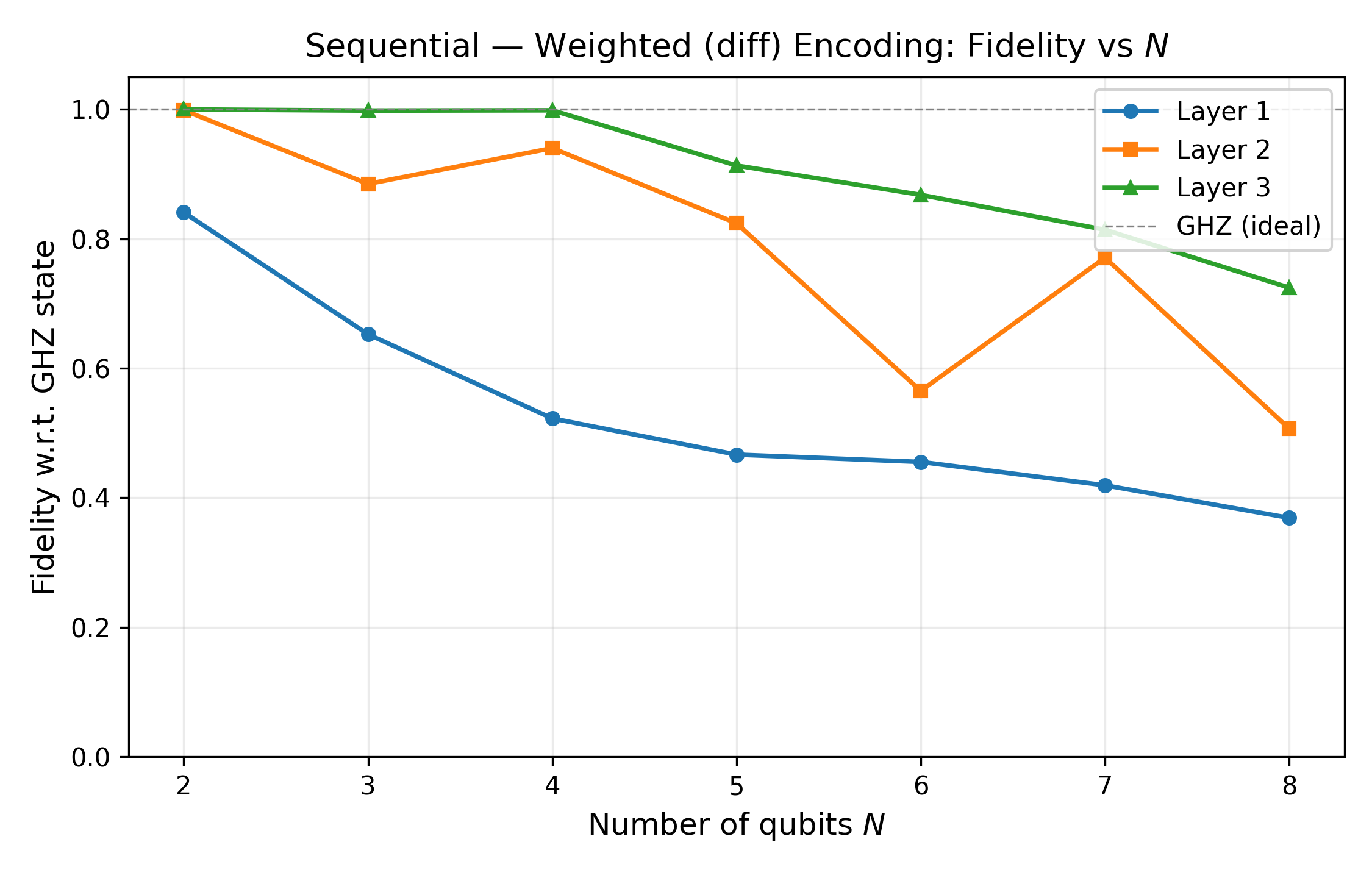}
    \caption{%
    GHZ fidelity versus $N$ for the sequential encoder, weighted-central
    encoding $\vec{\alpha} = (1, 0.5, \ldots, 0.5)$. Layout as in
    Fig.~\ref{fig:fid_uniform}.
    }
    \label{fig:fid_weighted}
\end{figure}

\paragraph*{Weighted-central encoding.}
Figure~\ref{fig:fid_weighted} shows the corresponding fidelity for $\vec{\alpha} = (1, 0.5, \ldots, 0.5)$. The overall pattern is similar: fidelity improves with depth and generally decreases with $N$. At $L = 3$, the encoder achieves fidelities above $0.91$ for $N = 2$--$5$ and $0.87$
at $N = 6$, comparable to or slightly higher than the uniform case at the same system sizes. At $N = 7$--$8$, the weighted fidelity at $L = 3$
reaches $0.83$ and $0.73$, respectively. Although the GHZ state is not the exact optimum for the asymmetric generator, these values show that the
encoder still finds states with substantial GHZ-like character under weighted encoding.
Notably, the $L = 2$ weighted fidelity at $N = 6$ (${\sim}0.57$) is markedly lower than the uniform value at the same point (${\sim}0.81$), reflecting the additional optimization difficulty imposed by the asymmetric generator at intermediate depth. The $L = 2$ curve also exhibits a non-monotonic uptick from $N = 6$ to $N = 7$ (${\sim}0.77$), suggesting
that the optimizer finds a qualitatively different solution at $N = 7$ that partially recovers GHZ-like character despite the larger system size. By $L = 3$ this irregularity is absent and the fidelity decreases more smoothly with $N$.

\subsection{Discussion}
\label{subsec:discussion}

We now interpret the decoder gains and fidelity results in light of the measurement-basis considerations introduced in Sec.~\ref{subsec:decoder}.

\paragraph*{Two-effect framework.}
The decoder gains reported in Sec.~\ref{subsec:sequential} are consistent with a two-effect picture. The first effect, convergence compensation, accounts for gains that arise whenever the encoder has not fully reached the target state, and the fixed Ramsey readout is therefore suboptimal. This effect is present in both encodings and is generally reduced at greater circuit depth as the encoder improves. The uniform data support this interpretation. At $N = 8$, the decoder gain falls from $6.2\%$ at $L = 1$ to $3.0\%$ at $L = 3$ (though non-monotonically, peaking at $7.5\%$ at $L = 2$), alongside a simultaneous increase in GHZ fidelity from $0.37$ to $0.69$ (Fig.~\ref{fig:fid_uniform}). For $N \leq 4$, where the encoder essentially saturates the HL by $L = 3$, the decoder gain is negligible at all depths. More broadly, configurations with GHZ fidelities above $0.97$ (uniform $N = 2$--$5$ at $L = 3$) show decoder gains well below $0.1\%$, indicating that fixed Ramsey readout is near-optimal once the probe state is sufficiently well prepared.

The second effect, structural basis misalignment, appears specifically under weighted encoding, where the asymmetric generator breaks the permutation symmetry of the polygon geometry and can shift the optimal measurement basis away from the standard Ramsey axis. Unlike convergence compensation, this effect can persist even as the encoder improves. At $N = 8$, the weighted decoder gain increases from $4.9\%$ at $L = 1$ to $7.0\%$ at $L = 3$, despite the GHZ fidelity rising from $0.37$ to $0.73$ over the same range (Fig.~\ref{fig:fid_weighted}). Though modest in absolute terms, this decoupling of fidelity improvement from decoder-gain reduction is consistent with a structural mismatch, in which the probe state improves but the optimal readout for the weighted generator remains distinct from the standard Ramsey basis. This persistent component of the decoder gain under weighted encoding raises the question of whether the ansatz itself can produce the entanglement structure that the asymmetric generator demands, independent of the measurement basis. We note that this two-effect picture is a data-driven interpretation consistent with the numerical evidence, not a formal decomposition of the decoder's contribution.

\paragraph*{Geometry-ansatz limitation.}
Beyond the decoder analysis, the weighted-encoding results show a second feature specific to the ansatz. The weighted-encoding CFI fails to sustain monotonic growth with $N$ at large system sizes, a behavior not observed under uniform encoding at the same depths. At $L = 3$, the encoder-only CFI peaks at $N = 6$ and drops at $N = 7$--$8$, with a slight recovery at $N = 8$ that does not restore the peak. At $L = 1$, the peak likewise occurs at $N = 6$ with a steady decline thereafter. The $L = 2$ landscape is more complex. The CFI drops from $N = 4$ through $N = 6$, recovers at $N = 7$ to the highest value across all $N$, and falls sharply at $N = 8$, indicating that the optimizer finds a qualitatively different solution at $N = 7$ for this depth. This irregular behavior persists across all seed runs at every depth, indicating that it reflects a structural feature of the optimization landscape rather than seed-dependent noise.

The polygon-centered geometry was chosen because it naturally distinguishes the central site from the peripherals, matching the structure of the weighted-central encoding. Nevertheless, as the peripheral shell grows, the ansatz appears to become progressively less expressive for the asymmetric entanglement structure demanded by the weighted generator. Notably, the GHZ fidelities at $N = 7$--$8$ are comparable between the two encodings (both reaching ${\sim}0.7$--$0.8$ at $L = 3$), yet only the weighted-encoding CFI exhibits these growth irregularities. This comparison constrains the possible explanations. Insufficient entanglement generation is unlikely given the comparable GHZ fidelities, and seed-dependent optimizer failure is unlikely given the consistency across seed runs. A measurement-basis mismatch alone is also insufficient, since the decoder gains remain bounded. The residual explanation within the present ansatz family is a mismatch between the entanglement produced by the polygon-symmetric ansatz and the asymmetric entanglement required by the weighted generator, motivating exploration of alternative lattice topologies in future work.

\paragraph*{Multi-seed methodology.}
The multi-seed CMA-ES approach is essential for reliable optimization at large system sizes. Seed variance data (Appendix~\ref{app:seeds}) show spreads of up to $48\%$ between best and worst seeds at $N = 7$--$8$ for $L = 1$, confirming that the CFI landscape is highly multimodal. Approaches using fewer or closely spaced seeds can exhibit non-monotonic dips that reflect optimizer variance rather than physical behavior. The 5-seed approach with well-separated seeds yields monotonic CFI scaling with $N$ at each depth for uniform encoding. For $L = 3$ at $N = 6$--$8$, where the landscape is most challenging, 4 additional seeds were used (9 total). The resulting spreads of $34$--$51\%$ at these configurations show that even 5 seeds may be insufficient at larger system sizes and deeper circuits.

Taken together, the encoding-dependent decoder gains, the geometry-ansatz mismatch under asymmetric encoding, and the multimodal optimization landscape indicate that structured encoding introduces nontrivial probe-and-measurement optimization challenges that are not captured by the generator spectrum or the unconstrained precision bounds alone, even in formally single-parameter estimation settings.

%
%
\section{Conclusion}
\label{sec:conclusion}
We presented a variational framework for structured linear function estimation in a dipolar spin-$1/2$ array, optimizing entangled probe states with respect to the CFI using CMA-ES with a multi-seed strategy. A shallow global decoder $V(\alpha,\beta)$, derived from the Euler decomposition with the trailing $R_Z$ shown redundant under $Z$-basis measurement, is optimized independently to assess whether the standard Ramsey readout extracts all available information. For both uniform and weighted-central encodings with $N = 2$--$8$ qubits and depths $L = 1$--$3$, the optimized probes approach the respective EE precision bounds, with the decoder gains physically interpretable as convergence compensation (diminishing with depth) and structural basis misalignment (persisting under asymmetric encoding). Under weighted encoding at large $N$, the CFI additionally exhibits non-monotonic growth across depths and seed runs, consistent with an expressivity limit of the polygon-symmetric ansatz rather than with optimizer or measurement failure.

These results demonstrate that hardware-motivated variational circuits can prepare near-optimal probe states for structured parameter estimation under finite-depth constraints, and that a simple diagnostic decoder can quantify the adequacy of the standard measurement protocol.

Several directions remain open. Incorporating realistic noise sources is essential for assessing practical relevance. Under noise, the decoder may play a more prominent role by compensating for noise-induced measurement-basis shifts, motivating noise-aware variational training. Joint probe-measurement co-optimization, in which all encoder and decoder parameters are trained simultaneously, may reach states unavailable to the sequential pipeline and warrants systematic investigation. Extension to multiparameter estimation is a natural next step that would exploit the full generality of the framework developed here. Alternative lattice topologies that better match the symmetry of asymmetric generators could alleviate the geometry-ansatz mismatch identified at large $N$, and benchmarking at larger system sizes will be necessary to assess scalability beyond the small-$N$ regime studied here.
\begin{acknowledgments}
The authors acknowledge support from the Pitt Momentum Fund. This research was also supported in part by the University of Pittsburgh Center for Research Computing and Data  (RRID: SCR\_022735) through the computational resources provided. Specifically, the work utilized the HTC and H2P clusters, which are supported by NIH award number S10OD028483 and NSF award number OAC-2117681. KPS thanks the U.S. Department of Energy, Office of Science, Advanced Scientific Computing Research (ASCR) program, for support under Award Number DE-SC0026264, and PQI Community Collaboration Awards.
GD also acknowledges support from NSF Award No.~2304998.


\end{acknowledgments}

\bibliographystyle{apsrev4-2}
\bibliography{bibliography}
\clearpage

\appendix

%
%
\label{app:bounds}
\section{Precision Bounds for Directional Encoding}
\label{app:bounds}

This appendix states the Standard Quantum Limit (SQL) and 
entanglement-enhanced (EE) precision bound for estimating the scalar 
quantity $q = \vec{\alpha}\!\cdot\!\vec{\phi}$, with local phases $\phi_i = \alpha_i\theta$. The directional encoding collapses the problem 
to a single-parameter estimate with a known structure set by the weight 
vector $\vec{\alpha}$. Following the analysis of Eldredge 
\textit{et al.} ~\cite{PhysRevA.97.042337} and assuming a fixed interrogation time $t = 1$, we quote the resulting bounds that apply to separable versus entangled probe states.

Let $\vec{\alpha} = (\alpha_1, \dots, \alpha_N)$. Defining 
$\|\vec{\alpha}\|^{2} = \sum_i \alpha_i^{2}$ and $S = \sum_i |\alpha_i|$, the effective parameter is $q = \|\vec{\alpha}\|^{2}\,\theta$. For optimally entangled probes, the minimal achievable variance is
\begin{equation}
\operatorname{Var}(q) \ge \frac{\|\vec{\alpha}\|^{4}}{S^{2}}
\quad\Longrightarrow\quad
\mathcal{F}_{\text{EE}}(q) = \frac{S^{2}}{\|\vec{\alpha}\|^{4}},
\end{equation}
while for fully separable probes measured locally,
\begin{equation}
\operatorname{Var}(q) \ge \|\vec{\alpha}\|^{2}
\quad\Longrightarrow\quad
\mathcal{F}_{\text{SQL}}(q) = \frac{1}{\|\vec{\alpha}\|^{2}}.
\end{equation}

For uniform encoding ($\alpha_i = 1/N$), $S = 1$ and $\|\vec{\alpha}\|^{2} = 1/N$, giving $\mathcal{F}_{\text{SQL}} = N$ and 
$\mathcal{F}_{\text{EE}} = N^{2}$ (the Heisenberg limit). For weighted-central encoding ($\alpha_c = 1$, $\alpha_i = 0.5$ for $i \neq c$), $S = (N{+}1)/2$ and $\|\vec{\alpha}\|^{2} = (N{+}3)/4$. The resulting SQL and EE values for 
both encodings are collected in Table ~\ref{tab:bounds_comparison}; these 
serve as reference targets for the variational optimization results in the main text.

\begin{table}[htbp]
\centering
\caption{SQL and EE bounds for uniform and weighted-central encodings.}
\label{tab:bounds_comparison}
\begin{tabular}{c cc cc}
\toprule
& \multicolumn{2}{c}{Uniform} & \multicolumn{2}{c}{Weighted-central} \\
\cmidrule(lr){2-3} \cmidrule(lr){4-5}
$N$ & SQL & EE & SQL & EE \\
\midrule
2 & 2.000 & 4.000  & 0.800 & 1.440 \\
3 & 3.000 & 9.000  & 0.667 & 1.778 \\
4 & 4.000 & 16.000 & 0.571 & 2.041 \\
5 & 5.000 & 25.000 & 0.500 & 2.250 \\
6 & 6.000 & 36.000 & 0.444 & 2.420 \\
7 & 7.000 & 49.000 & 0.400 & 2.560 \\
8 & 8.000 & 64.000 & 0.364 & 2.678 \\
\bottomrule
\end{tabular}
\end{table}

%
%
%
\label{app:trotter}
\section{Trotterization Stability}
\label{app:trotter}

The dipolar interaction blocks $\exp(-it\hat{H}_{\text{int}})$ in the 
variational ansatz are implemented via first-order Trotterization with 
$m$ steps per block. To verify that the reported CFI values are 
insensitive to the choice of $m$, we re-evaluated the CFI of all 42 
optimized configurations (both encodings, $N = 2$--$8$, $L = 1$--$3$) at 
Trotter step counts $m = 25$, $50$, $100$, $200$, $800$, and $1600$, 
using the production-optimized encoder and decoder parameters throughout. 
No re-optimization was performed; only the Trotter discretization was 
varied.

\begin{figure*}[htbp]
    \centering
    \includegraphics[width=\textwidth]{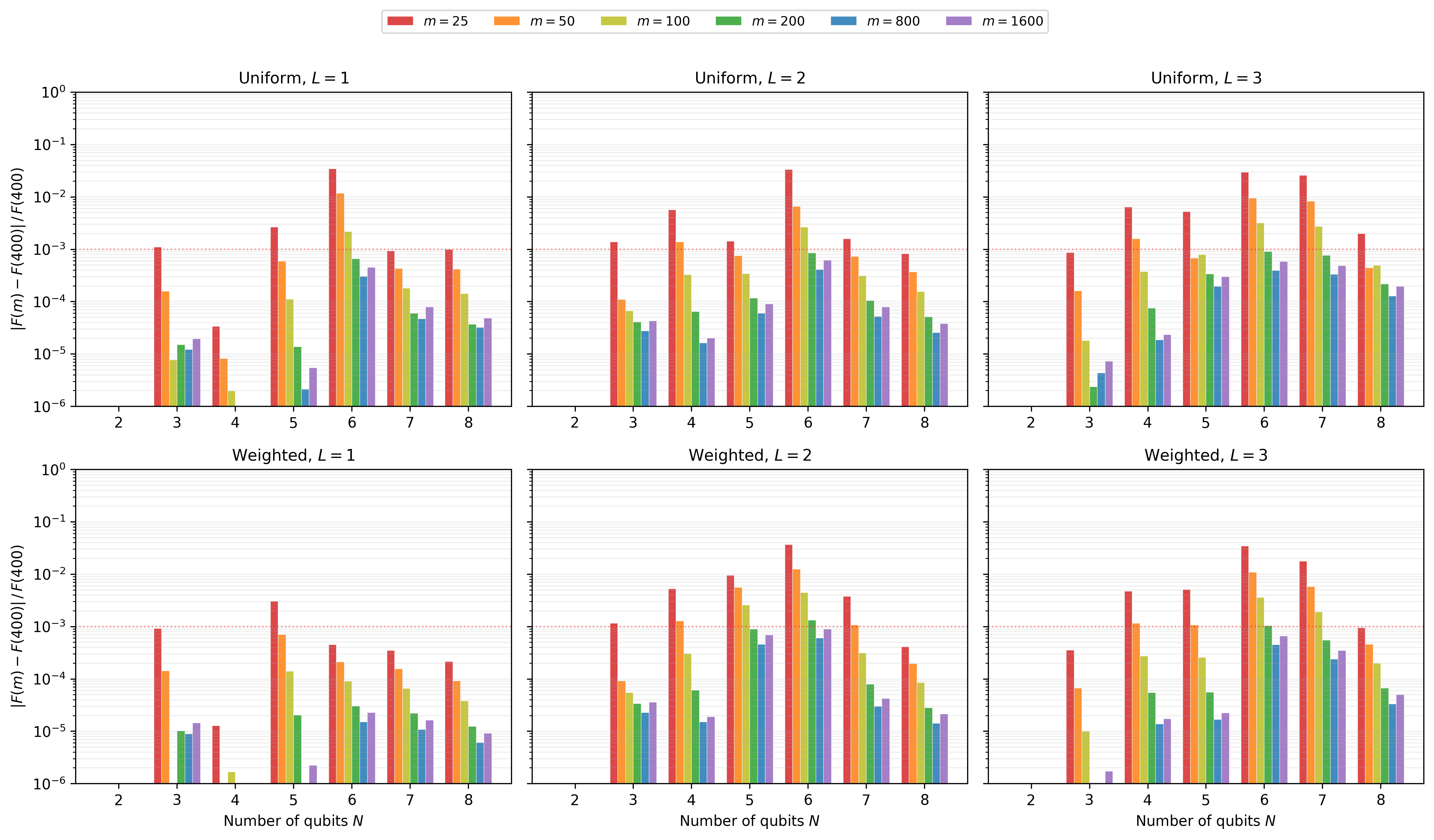}
    \caption{%
    Trotterization convergence: relative deviation 
    $|F(m) - F(400)| / F(400)$ of the CFI at Trotter step counts 
    $m = 25$, $50$, $100$, $200$, $800$, and $1600$ from the production 
    value $m = 400$, for all $(N, L)$ configurations under uniform (top 
    row) and weighted-central (bottom row) encodings. The red dashed line 
    marks $10^{-3}$. Bars for $N = 2$ are below $10^{-6}$ and fall 
    outside the plotted range.
    }
    \label{fig:trotter}
\end{figure*}

Figure~\ref{fig:trotter} shows the relative deviation 
$|F(m) - F(400)| / F(400)$ for each configuration, with the production 
value $m = 400$ as the reference. At coarse discretizations 
($m = 25$--$50$), the deviation can reach a few percent for the 
larger and deeper circuits, confirming that $m = 400$ is necessary for 
accurate results. At finer discretizations ($m = 800$, $1600$), the 
deviation falls below $10^{-3}$ for all configurations and below 
$10^{-4}$ for the majority, confirming that $m = 400$ is also sufficient 
and that increasing the Trotter count further would not meaningfully 
change the reported CFI values.

%
%
%
\label{app:cmaes}
\section{CMA-ES Hyperparameters}
\label{app:cmaes}

The sequential optimization pipeline consists of three stages: encoder 
coarse search, encoder local refinement, and decoder optimization. 
Table~\ref{tab:hyperparams} summarizes the CMA-ES settings common across 
system sizes, and Table~\ref{tab:perqubit} gives the per-qubit population 
size, step size, and timeout that scale with system size.

The encoder coarse stage uses 5 independent seeds with well-separated 
values for all configurations, with 4 additional seeds (9 total) for 
$N = 6$--$8$ at $L = 3$ where the landscape is most multimodal. The 
refinement stage runs a single seed initialized from the best coarse 
result with a tight step size for local search; if the refined CFI does 
not exceed the coarse CFI, the coarse parameters are retained, ensuring 
monotonic improvement. The decoder stage uses 3 seeds, with the fixed 
Ramsey parameters retained if no seed improves upon the encoder-only 
baseline.

\begin{table}[htbp]
\centering
\caption{CMA-ES settings common across system sizes for each pipeline 
stage.}
\label{tab:hyperparams}
\begin{tabular}{l l l l}
\toprule
Parameter & Enc.\ coarse & Enc.\ refine & Decoder \\
\midrule
Seeds & \{204, 604, & \{204\} & \{204, 604, \\
      & 1204, 2004, & & 1204\} \\
      & 3004\}$^{a}$ & & \\
$\sigma_0$ (global) & Table~\ref{tab:perqubit} & 0.20 & 0.55 \\
\texttt{CMA\_stds} & $= \sigma_0$ & 0.15 & --- \\
\texttt{tolstagnation} & $50{+}10(L{+}N)$ & 120 & 120 \\
Bounds & $[0,1]^{3L\,b}$ & $[0,1]^{3L\,b}$ & unbounded \\
Init.\ jitter & --- & $10^{-3}$ & --- \\
Warm start & from $L{-}1$ & from coarse & from $(0,\pi/2)$ \\
\bottomrule
\multicolumn{4}{l}{\footnotesize $^{a}$Plus \{4004, 5004, 6004, 7004\} 
for $N = 6$--$8$ at $L = 3$.} \\
\multicolumn{4}{l}{\footnotesize $^{b}$Normalized: $t_1, t_3 \in [0,1]$,
$\theta_2 \in [-1,1]$.}
\end{tabular}
\end{table}

\begin{table}[htbp]
\centering
\caption{Per-qubit CMA-ES population size, initial step size, and timeout 
for each pipeline stage.}
\label{tab:perqubit}
\begin{tabular}{c ccc ccc cc}
\toprule
& \multicolumn{3}{c}{Encoder coarse} 
& \multicolumn{3}{c}{Encoder refine}
& \multicolumn{2}{c}{Decoder} \\
\cmidrule(lr){2-4} \cmidrule(lr){5-7} \cmidrule(lr){8-9}
$N$ & Pop & $\sigma_0$ & Time 
    & Pop & $\sigma_0$ & Time 
    & Pop & Time \\
\midrule
2--3 & 30 & 0.45 & 4\,h & 28 & 0.15 & 3\,h & 24 & 8\,h \\
4--5 & 40 & 0.45 & 6\,h & 28 & 0.15 & 5\,h & 24 & 8\,h \\
6    & 50 & 0.45 & 8\,h & 30 & 0.15 & 6\,h & 24 & 8\,h \\
7    & 60 & 0.50 & 10\,h & 32 & 0.15 & 7\,h & 24 & 8\,h \\
8    & 70 & 0.50 & 12\,h & 34 & 0.15 & 8\,h & 24 & 8\,h \\
\bottomrule
\end{tabular}
\end{table}

All circuit evaluations use $m = 400$ first-order Trotter steps 
(convergence verified in Appendix~\ref{app:trotter}) and a probability 
cutoff $p_{\min} = 10^{-12}$ below which outcome probabilities are 
excluded from the CFI sum. The CFI is evaluated at $\theta = \pi/100$ 
using the exact parameter-shift rule with shift $s = \pi/2$, requiring 
$(2N{+}1)$ circuit evaluations per fitness call. For the layerwise warm 
start, depth $L = 1$ is initialized from a broad random distribution 
(time parameters from $\mathrm{Unif}(0.1, 1.0)$, angle parameters from 
$\mathrm{Unif}(0.0, 1.0)$ in normalized units); for $L > 1$, the best refined parameters from $L{-}1$ are extended with a new layer initialized 
from the same distribution.

%
%
\label{app:cfi_tables}
\section{Seed Variance}
\label{app:seeds}

\begin{figure*}[htbp]
    \centering
    \includegraphics[width=\textwidth]{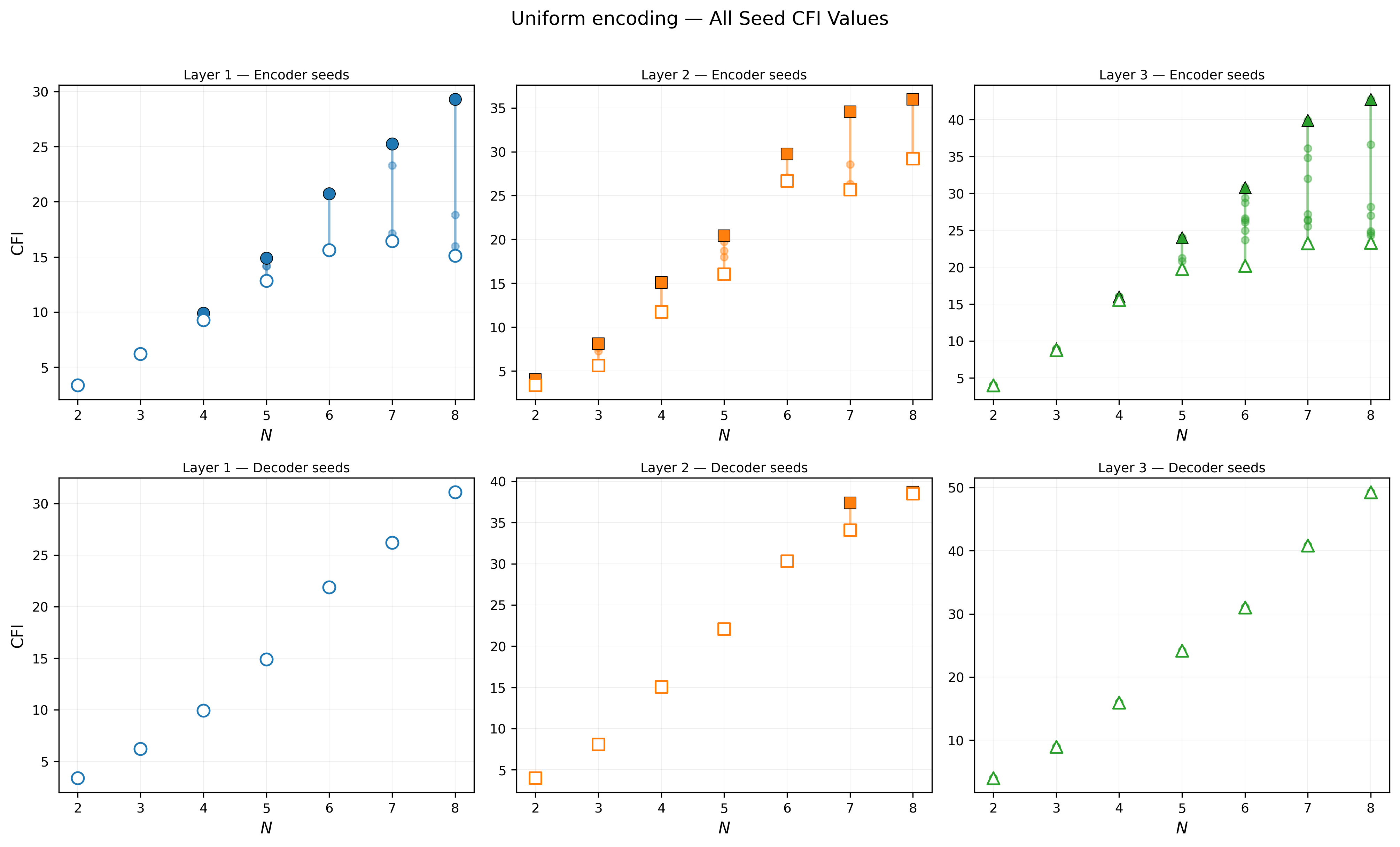}\\[4pt]
    \includegraphics[width=\textwidth]{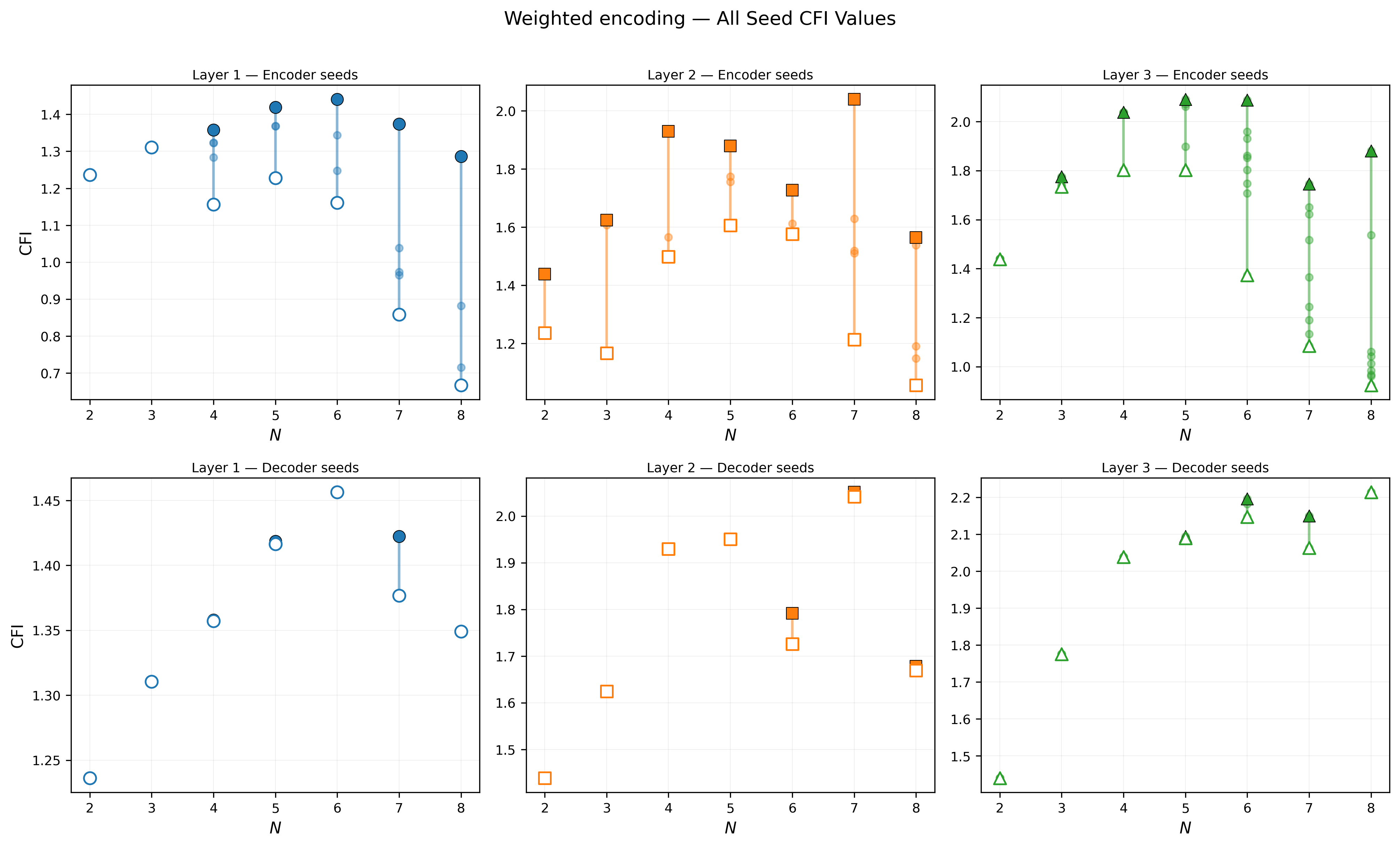}
    \caption{%
    Seed variance for uniform encoding (top) and weighted-central encoding 
    (bottom). Each panel shows the CFI achieved by individual seeds for 
    the encoder (upper subpanels) and decoder (lower subpanels) stages at 
    $L = 1$, $2$, and $3$. Filled markers indicate the best seed retained 
    for subsequent stages.
    }
    \label{fig:seeds}
\end{figure*}

Figure~\ref{fig:seeds} shows the CFI achieved by each individual seed 
for both the encoder and decoder stages across all $(N, L)$ 
configurations. Filled markers denote the best seed retained for 
subsequent pipeline stages; open markers show remaining seeds. The 
vertical spread at each $N$ directly visualizes the landscape 
multimodality: for small $N$ the seeds collapse to a single point, while 
for $N \geq 6$ the spread is substantial, reaching $34$--$51\%$ at 
$L = 3$. At $L = 3$ for $N = 6$--$8$, the denser clusters reflect the 4 
additional seeds (9 total) used at these system sizes, with the best seed 
sometimes found among the additional seeds (e.g., seed 6004 for uniform 
$N = 7$, $L = 3$).

\end{document}